\documentclass[aps]{revtex4-1} 

\usepackage{graphicx}
\usepackage{color}
\usepackage{epsfig} 
\usepackage{pstricks}

\begin{document}

\title{Hybrid density-functional theory calculations on surface effects in Co doped ZnO nanowires}

\author{A. L. Rosa}
\affiliation{Instituto de F\'{\i}sica, Universidade Federal de Goi\'{a}s, 74.690-900 Goi\^{a}nia, Goi\'{a}s, Brazil}
\affiliation{Bremen Center for Computational Materials Science, University of Bremen, Am Fallturm 1, 28195, Bremen, Germany}
\author{L. L. Tacca}
\affiliation{Instituto de F\'{\i}sica, Universidade Federal de Goi\'{a}s, 74.690-900 Goi\^{a}nia, Goi\'{a}s, Brazil}
\author{Th. Frauenheim}
\affiliation{Bremen Center for Computational Materials Science, University of Bremen, Am Fallturm 1, 28195, Bremen, Germany}

\date{\today}

\begin{abstract}
  In this work we have employed density-functional theory with hybrid
  functionals to investigate the atomic and electronic structure of
  bare and hydrogenated Co doped ZnO nanowires. We find that in the
  absence of passivation on the nanowire surface, the cobalt atoms
  segregate to the surface. On the other hand, under hydrogen
  passivation, the incorporation of Co is more favorable at inner
  sites. This suggests that the incorporation of Co in nanostructures
  has a dependence on the environment and may be facilitated by
  external atoms and relaxation of the surface.
\end{abstract}

\maketitle
\section{Introduction}
\label{sec:s1} 

Zinc oxide (ZnO) has been largely investigated in the past years it is
a low cost, high electron mobility, it is transparent and can be
sinthesized in several nanostructured shapes\,\cite{ZLWang2004}. Doping
in ZnO has been widely used to tailor its electronic, magnetic and
optical properties.  In particular, cobalt doped ZnO nanostructures
have been largely investigated in the past years both
experimentally\,\cite{Wang:12,Castro2016,Wojnarowicz:2015,Nanomaterials2017,Chanda:17,Geburt:2013}
and theoretically \,\cite{Sarsari:13,Dalpian2006,Patterson,SciRep2017}
due to the its promising application in optoelectronics and
spintronics. In nanostructures, as the surface/volume ratio is large,
the influence of surface effects on the incorporation of impurities in
nanostructures can play an important role. Effects of surface passivation, morphology of the surface and co-doping can influence the incorporation of dopantes, as has been discussed in Ref.\cite{Erwin2005}. In this letter, we
investigate Co incorporation in ZnO nanowires in order to determine
changes in the magnetic properties and electronic structure upon
surface passivation. Our modeling mimics certain experimental
conduitions whjere air or hydrogen atmosphere is present. In general
local density functionals lead to wrong description of band gaps. In
order to reproduce experimental gap of ZnO and provide a better
description of we have performed density functional theory
calculations with hydbrid functionals. We show that although there is
a strong localization of the Co states with no significant change in
its magnetic moment. However there is a site preference depending on
the wire surface termination.

\section{Computational details} 

We have used density-functional theory (DFT)\,\cite{Kohn:65} together
with the projected augmented wave (PAW) method, as implemented in the
Vienna Ab initio Simulation Package (VASP)\,\cite{Kresse:99}. The
Heyd-Scuseria-Ernzenhof (HSE06) form of the exchange-correlation
potential was used to obtain geometries, formation energies and
magnetic moments. To model Co impurities in ZnO we built up a 96 atom
supercell using our calculated PBE lattice parameters of ZnO,
$a$=3.25{\AA} and $c$=5.25{\AA}. To ensure convergence of structural,
electronic and magnetic properties, a cutoff of 400 eV was used for
the plane-wave expansion of the wave function. Atomic forces were
converged up to 0.01\,eV/{\AA}. For Brillouin zone integrations, a
$(1\times 1 \times 4)$ Monkhorst-Pack {\bf k}-point sampling was used.  The
HSE results should provide a reliable method to determine the
electronic structure. Previous results have demonstrated that although
25\% of Hartree-Fock can be justified\,\cite{Lany:10}, to obtain the
experimental band gap of ZnO, 36\% HF is
needed\,\cite{Janotti:09,Lany:10}. Therefore, we have used this admixture to
reproduce the experimental gap of bulk ZnO\,\cite{CRC}. 
The predicted energy position of the minority spin
Co-t$2$ states is 3.0-3.6 eV above the ZnO conduction band minimum
which is closer to GW calculations\,\cite{Sarsari:13,ALRosaunpublished,Lorke:16}.

\section{Results}

\begin{figure}[ht!]
\includegraphics*[width=8cm]{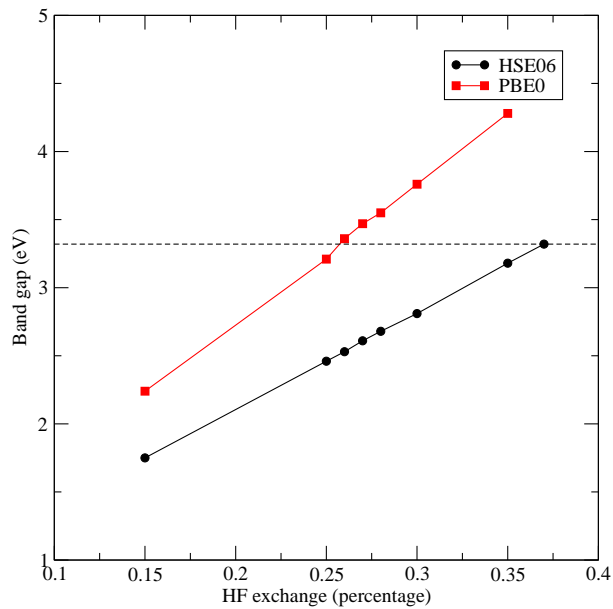}
\caption{\label{fig:benchmark} HSE and PBE0 energy gaps for ZnO as a function of the Hartree-Fock exchange admixture.}
\end{figure}

In Fig. \ref{fig:benchmark} we show the variation of the band gap of
ZnO bulk with the amount of Hartree-Fock added to the
exchange-correlation functional for both PBE0 and HSE. As reported
previously, due to the lack of screening, the PBE0 functional requires
a large amount of HF (around 40\,$\%$) to reach the experimental
gap\cite{Lany:10}. On the other hand the HSE functional can reproduce the
experimental gap with 25\,$\%$ admixture.

\begin{figure}[ht!]
\includegraphics*[width=8cm]{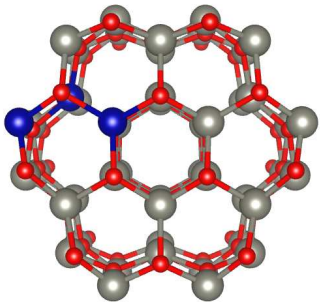}
\caption{\label{fig:geometries} Atomic configurations for ${\rm Co_{Zn}}$ in bare ZnO nanowires. O, Zn and Co atoms are shown in red, gray and blue color, respectively.}
\end{figure}

The geometry of the structures we have investigated is shown in
Figs.\,\ref{fig:geometries}. First we discuss the non-passivated
wires. A single Co occupying a substitutional Zn site in the middle of
the wire does not produce strong distortion in the ZnO wire lattice
(around 0.1\,{AA}). For Co sitting at subsurface/surface sites, Co
moves (outwards inwards), as the covalent radius is similar to
Zn. The Co-O bond lengths remain very close to the values in pure
ZnO, rangind from 1.8-2.1 \,{\AA} in bare wires and from 2.0-2.2
\,{\AA} in hydrogenated wires. The formation energy of an isolated
defect is calculated setting the energy zero to the minimal energy
configuration.

Next we discuss the thermodynamic stability of Co in these small
diameter wires. In the case of hydrogen adsorption, the most
stable configuration is a fully hydrogenated wire.  The incorporation
of a Co atom in the wires has a dependence with the site position. For
bare wires, the preferred position is the surface position. This
effect is called self healing because nanostructures have a small
volume compared to their surface, leading to the migration to the
surface\,\cite{Dalpian2006,darosa2010,Erwin2005}. It has been further
suggested size is not the only effect responsible for impurity
incorporation. Erwin et al.\,\cite{Erwin2005} suggested that impurity
incorporation depends on three main factors, surface morphology,
nanostructure shape, and surfactants. Indeed, in our previous work we
have shown that adsorption of hydrogen and water on the surface of ZnO
wires is an exothermic reaction
\cite{XuAPL2007,Fan:07,XuPRB2009}. Later on we have investigated the
incorporation of N in ZnO nanowires and showed the effect of hydrogen
passivation\,\cite{darosa2010} under N incorporation which has been
confirmed experimentally that N should sit close to surface sites and
at oxygen positions\,\cite{Buyanova2015}. We show here that a
different behavior is found for Co in ZnO ultrathin wires. By
passivating the nanowire surfaces with hydrogen, Co have a lower
formation energy than when it is incorporated in the bulk (for both
PBE and HSE functionals), as shown in Table\,\ref{table:formation}.
Specially for these ultrathin wires, this effect is dramatic for
passivation with hydrogen, since the energy difference between bulk
and surface position is 1.0 eV with HSE. Main factors for this
behavior is that Co has a similar size as ZnO.

\begin{figure}[t]%
\includegraphics*[width=8cm]{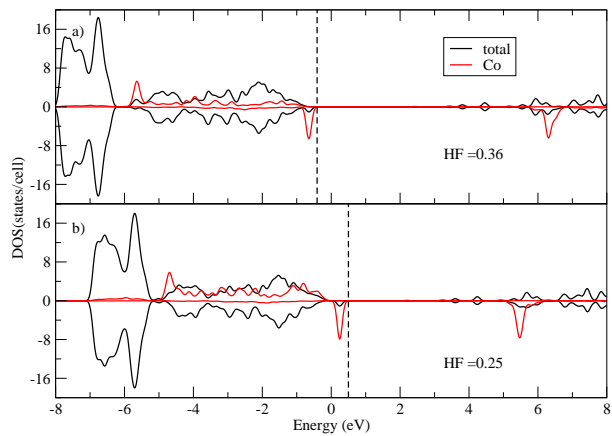}
\caption{Total and projected density-of-states for Co doped ZnO bulk with HSE functional. a) with 36$\%$ admixture and b) with 25$\%$ admixture. The Fermi level is represented as a dashed line).}
\label{fig:dos_bulk}
\end{figure}

\begin{table}[h]
  \caption{\label{table:formation} Total magnetic moments $\mu_{\rm tot}$ (in $\mu_{\rm B})$ and relative formation  energies $\rm E_f$ (in eV) of
  neutral Co impurities in ZnO calculated with PBE and HSE functionals. In brackets the magnetic moment projected on the Co atoms is shown.}
\begin{tabular}{lcccc}
\hline
\hline
config.      & \multicolumn{2}{c}{$\mu_{\rm tot}$} & \multicolumn{2}{c}{$\rm E_f$}\\
\hline
             &   PBE   &   HSE    &  PBE & HSE\\
bare  inner  &   3.10(2.50) & 3.00(2.69)     & 0.25    & 0.24\\
bare sub     &   3.15(2.53)   &    3.00 (2.70) &  0.20   &  0.78\\
bare surf    &   3.14(2.50)   &   3.00 (2.70)   &  0.00   &   0.00\\
\hline
hydro inner  &   3.00(2.45)    &   3.00 (2.70) &  0.00   &  0.00\\
hydro sub    &  3.00(2.46)    &    3.00 (2.70) &  0.01  & 0.21\\
hydro surf   &  3.00(2.34)    &  3.00 (2.65)  &   0.22  & 1.00\\
\hline
\hline          
\end{tabular}
\end{table}

\begin{figure}[ht!]
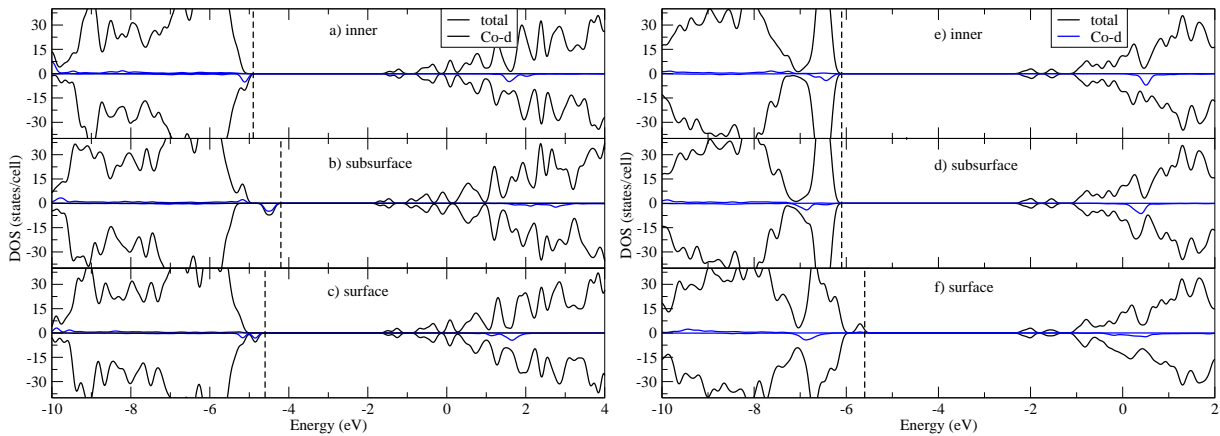

\includegraphics*[width=8cm]{./dos_cozno_nw_bare.eps} 
\includegraphics*[width=8cm]{./dos_cozno_nw_hydro.eps}
\caption{\label{fig:dos_doped} Density of states of Co doped ZnO nanowires. (a) - (c) bare wires and (d)-(f)  hydrogenated wires with HSE and 0.36 HF exchange. The vertical dashed line represents the Fermi level which is set to the highest occupied state.}
\end{figure}

Doping of ZnO bulk with Co splits the Co-3$d$ states into $e$ and
$t_{2}$ states. Co assumes a high-spin configuration with local
magnetic moment of 2.7 $\mu_{\rm B}$ with a small hybridization with
neighboring O atoms. The majority spin states $e$ are close to the
valence band maximum of ZnO and therefore hybridize with the O-2$p$
states. The $e$ minority spin states lie right above the ZnO valence
band maximum. The amount of Hartre-Fock exchange changes slightly the
position of the $e$ minority spin states. Including 0.25 of HF
exchange in the DFT exchange term places these states 0.5 eV above the
top of the valence band. The distance between the $e$ minority spin
states $e$ and the t$_{2}$ states is shown in Fig.\,\ref{fig:dos_bulk}
(a). We obtain a value of 5.3\,eV, in agreement with
Refs.\,\cite{Sarsari:13,Patterson}. By increasing the amount of HF
exchange to 0.36 the $e$ minority states are pushed towards the
valence band top. The $e$-t$_{2}$ energy difference increases to 7\,eV as
shown in Fig.\,\ref{fig:dos_bulk} (b).

In Fig.\ref{fig:dos_doped} the DOS for the doped wires is
shown. Figs.\ref{fig:dos_doped} (a)-(c) shows the results for bare
wire.  When the Co incorporation is at inner site
(Fig.\ref{fig:dos_doped} (a)) the location of the $e$ minority states
is similar to the one in ZnO bulk. As the Co moves towards the
surface, relaxation and symmetry breaking effects yield to Co $e$
states located slightly at higher energies inside the band gap, but
still similar to the DOS in Co doped ZnO bulk, as it can be seen in
Fig.\ref{fig:dos_doped} (b) and (c).  In bare wires, the distance between
the $e$ and $t_2$ states for different sites is 6.7 eV (inner), 6.5 eV
(subsurface) and 6.6 eV (surface).

The DOS in hydrogenated wires shows some noticeable differences. Now
the $e$ minority states lie inside the valence band and are shifted to
lower energies as the Co atom is positioned towards the surface. This
means that hybridization of cobalt with ZnO may be tuned by
hydrogenation or incorporation of other adsorbing species in the
sample during the doping process. One of the reasons is that strain at
surface sites may change due to adsorption, facilitating Co
incorporation. We have previously shown that strain indeed can induce
diffusion of vacancies towards the surface\,\cite{Deng2014,Kou2017}.  Obviously this is a
kinetic process and barriers for Co diffusion under this conditions
need to be investigated to confirm this idea. 
However, the strain does not affect considerably the distance between
the $e$ and $t_2$ states minority spin states, which are 7.0 eV
(inner), 6.3 eV (subsurface) and 6.3 eV (surface). 

\section{Conclusions}

We have investigated ZnO nanowires doped with Co using hybrid
functionals.  We show that the impurity prefers to sit at bulk
positions when the surface is passivated. On the other hand, bare
wires suffer from self-purification problems leading to segregation of
the dopant of Co towards surface sites. This indicates that the
impurity can be more easily incorporated depending on the atmosphere
it is prepared and may be facilitated by external atoms and relaxation
of the surface. A route to investigate the diffusion of such
impurities in nanostructures upon different enviroment conditions
would provide further insights on these complex systems.

\section{Acknowledgements}
We are thankful for the financial support from the Brazilian agencies
CNPq and FAPEG. A.L.R and T.F. would like to thank also German Science Foundation (DFG) under the program FOR1616.


\providecommand{\WileyBibTextsc}{}
\let\textsc\WileyBibTextsc
\providecommand{\othercit}{}
\providecommand{\jr}[1]{#1}
\providecommand{\etal}{~et~al.}

\end{document}